# Electronic reconstruction at the interface between band insulating oxides: the LaAlO$_3$/SrTiO$_3$ system

M. Salluzzo

CNR-SPIN, Complesso MonteSantangelo via Cinthia, 80126 Napoli, Italy
marco.salluzzo@spin.cnr.it

**Abstract** The conducting quasi-two dimensional electron system (q2DES) formed at the interface between LaAlO$_3$ and SrTiO$_3$ band insulators is confronting the condensed matter physics community with new paradigms. While the mechanism for the formation of the q2DES is debated, new conducting interfaces have been discovered paving the way to possible applications in electronics, spintronics and optoelectronics. This chapter is an overview of the research on the LAO/STO system, presenting some of the most important results obtained in the last decade to clarify the mechanism of formation of the q2DES at the oxide interfaces and its peculiar electronic properties as compared to semiconducting 2D-electron gas.

# 1. Introduction

Transition metal oxides (TMO) have been widely studied in the last decades for the wide range of electronic properties and related intriguing physical phenomena that characterize their phase diagrams, including colossal magnetoresistance in manganites [1] and high $T_c$ superconductivity in cuprates [2].

While a complete understanding of the extraordinary physics of these materials remains elusive, the theoretical descriptions of their bulk properties are becoming clearer and capable of explaining part of the phenomenology. However, the latest advancements in the atomic control of epitaxial TMO heterostructures are confronting the oxide community with new challenges [3]. Nowadays, ultra-thin TMO layers can be combined in artificial heterostructures characterized by interfaces with perfection at the atomic scales level [4]. Many laboratories around the world are now able to fabricate complex multi-layered oxides using conventional physical vapour deposition (PVD) techniques. The physical properties of these novel heterostructures may differ substantially from the bulk properties of the constituent layers. This is related to the distinctive and quite general characteristics of transition metal oxides, and in particular to the important role of electron and magnetic correlations for electrons in 3d bands [5]. TMO's are characterized by bulk carrier densities in the range of $10^{17}$ to $10^{21}$ carriers/cm$^3$, so that electron-electron on-site and inter-sites Coulomb repulsion is not (completely) screened. As a consequence the TMO's physics depends on interactions taking place on relative short characteristics lengths, of the order of few interatomic distances. For these reasons, the realization of TMO's heterostructures, with properties entirely dominated by the interface physics, required important technological improvements in the atomic control of the structural and chemical properties of each layer.

The quasi two-dimensional electron system (q2DES) discovered at the interface between insulating LaAlO$_3$ (LAO) thin films and bulk SrTiO$_3$ (STO) [6], the subject of this review, stands as a model of oxide heterostructures characterized by electronic properties uniquely determined by the interface. Interface-physics played a central role in the development of modern semiconductor based microelectronic, and determined the success of Metal-Oxide/Semiconductor field effect transistors (MOSFET). Noteworthy, while the bulk physics of semiconductors was well understood since long time, studies on confined electron gas in GaAs/AlGaAs quantum-wells and quantum dots were the playground for new discoveries. From this point of view, the observation of novel physical phenomena in TMO heterostructures characterized by atomic abrupt interfaces was not at all unexpected. However, the discovery by Ohtomo and Hwang of a 2D-conductivity at the interface between two insulating oxides came as a surprise in view of the large band gaps of both LaAlO$_3$ (5.6 eV) and SrTiO$_3$ (3.2 eV).

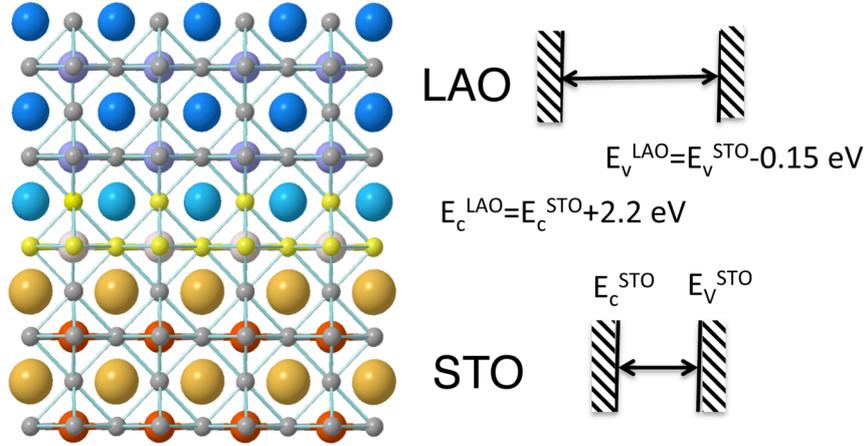

**Fig.1** Ideal LaAlO$_3$/SrTiO$_3$ interfacial layers. On the right, the bulk electronic configuration of LAO and STO conduction and valence bands.

The LaAlO$_3$/SrTiO$_3$ interface host a q2DES characterized by unique electronic properties, including respectable mobility exceeding 5x10$^4$ cm$^2$ V$^{-1}$ s$^{-1}$ (at 4.2 K) [7], low temperature superconductivity [8], and widely tuneable electric properties using electric field effect, which allows, for example, a control of metal to insulating transition [9-10] (even at room temperature) and a modulation of Rashba spin-orbit coupling over a large range [11].

This chapter is an overview of the state of art of research and future trends in the study of LaAlO$_3$/SrTiO$_3$ and related heterostructures. The main aim is to present well-established results as well as main debated controversies in a general perspective.

The chapter is organized as follows: Section 2 is devoted to the general properties of LAO/STO heterostructures, growth methods and known influence of deposition parameters. Section 3 is a concise overview of the mechanisms proposed to explain the microscopic origin of the metallic conductivity at the LAO/STO interface. Section 4 is dedicated to the main electronic and magnetic properties deduced from spectroscopic experiments with a comparison to the predictions of theoretical models. Section 5 will be a short summary and a discussion about future works and perspectives.

The chapter does not pretend to give a complete survey of the subject, thus any missing references to all relevant researches on the topic was unintentional.

# 2 The LaAlO$_3$/SrTiO$_3$ Interface

## 2.1 General properties of LaAlO$_3$ and SrTiO$_3$

LaAlO$_3$ and SrTiO$_3$ compounds are well-known band insulating oxides with perovskite ABO$_3$ structure. LaAlO$_3$ is well known as functional dielectric in microwave superconducting resonators, filters and antennae, due to its relatively low losses at microwave frequencies [12] and high dielectric constant ($\varepsilon_r$=24) among other oxides. SrTiO$_3$ is extremely popular in the oxide community as single crystalline substrate for the growth of transition metal oxide films; in particular it is widely used to fabricate high quality epitaxial manganites [13], high-T$_c$ cuprates [14, 15], and nichelate thin films [16]. SrTiO$_3$ is also known to be a quantum paraelectric material [17], i.e. a material for which quantum fluctuations prevent the onset of a ferroelectric long-range order; it is characterized by a diverging dielectric constant at low-T reaching values exceeding $\varepsilon_r$=20000. As matter of fact, SrTiO$_3$, as other quantum paraelectric materials, becomes ferroelectric under strain [18].

Concerning the band structure, as shown in Fig.1, the band gaps of these perovskites are 5.6 eV in the case of LAO [19] and 3.2 eV in the case of STO [20]. The valence bands of the two bulk compounds are very close to each others (shifted by about 0.15 eV), and are related to the overlapping of oxygen 2p (O2p) states partially hybridised with A and B cation orbitals. The conduction band minimum (CBM) of STO and LAO are shifted one from the other by about +2.2 eV [see Fig.1]. Indeed, stoichiometric STO is characterized by empty Ti-3d orbitals (with some admixture of O2p states), which are located close to the Fermi level. On the other hand, La and Al derived bands are quite far from the Fermi Level (of the order of 2.5 eV above the Fermi level). SrTiO$_3$ can be easily turned to a semiconducting and to a metallic state, with the peculiar characteristic, among other oxides, of requiring very small amount of chemical doping level to undergo the transition [21]. In particular, a bulk metallic character is achieved with much less than 0.1% of dopants. Moreover, STO is known to become superconducting at very low temperatures (200 mK-400 mK), and is characterized by a very peculiar temperature vs. doping phase diagram [22] characterized by a bell shape dome reminding those of High-Tc cuprates and pnictides. Superconductivity in STO is still not fully understood, since it occurs in a range of electron doping more than three orders of magnitude lower than classical BCS metallic superconductors [23].

From a technological point of view, a major advancement in the realization of atomic sharp interfaces employing STO was the control of the surface termination of (001) oriented STO single crystals [Fig.2a]. Kawasaki et al. [24] and later-on Koster et al. [25] used a combination of chemical etching and oxygen post annealing to realize perfectly single-terminated STO (001) substrates [Fig.2c]. Later studies demonstrated that these single crystals were TiO$_2$-terminated and

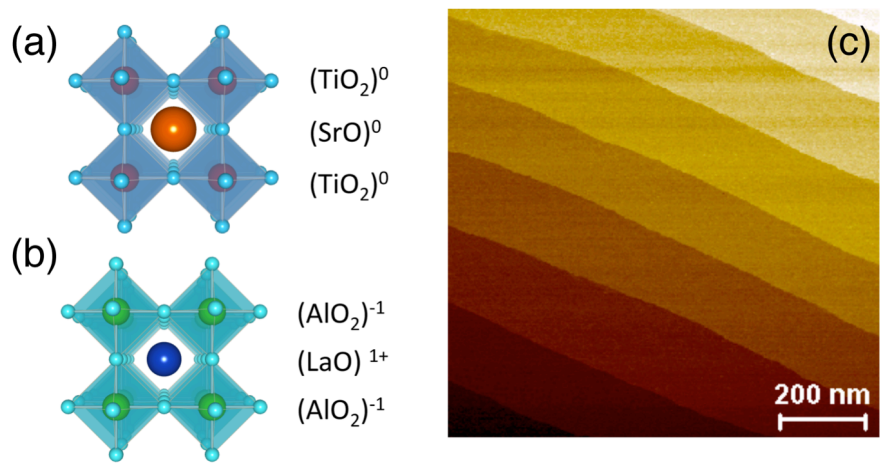

**Fig. 2.** The (a) SrTiO$_3$ and (b) LaAlO$_3$ unit cells, with the formal charge assignment to the (001) layers, and (c) typical atomic force microscopy image of the surface morphology of single TiO$_2$ terminated STO (001) obtained following the methods used in ref.[25].

structurally unreconstructed [26]; this is at odds with several previous surface studies on ultra high vacuum (UHV) prepared STO single crystals, which exhibited a wide range of reconstructions, like 2x1, 2x2, c2x2 and √5x √5-R26.6° [27], depending essentially on the annealing temperatures in UHV

The tendency towards non-stoichiometric reconstructed surfaces is a very interesting characteristic of STO (001) surfaces, and in some sense quite unexpected according to the common wisdom that both TiO$_2$ and SrO surfaces should be stable, being neutral. Actually, as discussed in some details by Vanderbilt et al. in ref. [28], SrTiO$_3$ is characterized by only partial ionic bonding, due to the charge transfer from oxygen to strontium and titanium cations. The degree of covalence is generally enhanced at the oxide surfaces and can change among the first atomic layers. Thus instead of having alternating, formally neutral, Sr$^{2+}$O$^{2-}$ and Ti$^{4+}$(O$_2$)$^{4-}$ planes, STO unit cells (uc) can be effectively weakly polar. This is extremely important for the surface and interface stability of heterostructures employing STO as substrate or as composing layer, an aspect often (almost always) overlooked.

LaAlO$_3$ (001) surface, on the other hand, has a strongly polar character [Fig.2b]. In particular, according to Tasker classification [29], AlO$_2$ and LaO layers are examples of type III surfaces. Each LAO unit cell shows a dipole moment that, without a surface structural reconstruction, causes an internal electrostatic potential diverging rapidly with the number of layers, a well know problem which have the name of "polar catastrophe instability". Thus the LAO (001) surfaces are intrinsically unstable and tend to have large deviations from the ideal flat termination. Very interestingly, some studies have shown that the most stable termination of LaAlO$_3$ single crystals is AlO$_x$ (x<2), with the Al coordination decreasing from octahedral to tetrahedral. Strong relaxation of cations toward the bulk and of the oxygen ions toward vacuum takes place at high temperatures [30].

Concerning ambient and low temperature bulk structures of $SrTiO_3$ and $LaAlO_3$, they are quite different one from the other, but close to ideal perovskites. At 300 K $SrTiO_3$ is cubic with a =3.905 Å, and shows a series of transitions at low temperatures and in particular a cubic to tetragonal transition around 108 K, related to rotations of $TiO_6$ octahedra. $LaAlO_3$ is known to be rhomboedric also at room temperature [31], while at high temperature is cubic. The pseudo-cubic lattice-parameter of $LaAlO_3$ is 3.79 Å. The mismatch of $LaAlO_3$ films respect $SrTiO_3$ is about 3%, which is not small. Indeed, thick $LaAlO_3$ films on STO (thickness above 15-20 nm) usually exhibit, together with twinning domains, cracks related to linear misfit dislocations.

To summarize, the growth of $LaAlO_3$ on $SrTiO_3$ presents many possible sources of instabilities that have to be eliminated by compensating mechanisms, more in details:
1) Valence and in particular conduction band mismatch at the interface;
2) Polarity discontinuity due to the different charges of the atomic layers;
3) Structural mismatch of LAO on STO, with in-plane compressive strain in the case of pseudomorphic growth;
4) Different octahedral rotations of the $BO_6$ units in the LAO and STO bulks;
5) Finally, and probably the most important, a polar instability of the LAO film due to its polar character. To less degree, also the STO (001) surface is unstable, being weakly polar.

All these instabilities certainly play a role in determining the characteristics of the system, from a structural, chemical and electronic point of view.

## 2.2 The growth of $LaAlO_3/SrTiO_3$ interfaces

In spite of the relatively large mismatch, $LaAlO_3$ is characterized by a pseudomorphic growth on $SrTiO_3$ (001). Pulsed Laser Deposition (PLD) assisted by Reflection High Energy Electron Diffraction (RHEED) is commonly employed to growth $LaAlO_3/SrTiO_3$ interfaces [32, 33], using single crystalline or ceramic $SrTiO_3$ and $LaAlO_3$ targets. PLD growth is very easy and allows, in principle, a quasi-stoichiometric transfer of the target composition to the substrate. On the other hand, it has been shown that even in the case of STO homoepitaxy, the overall stoichiometry, thus the defect density, of PLD grown films is extremely sensitive to the deposition conditions, in particular to the target to substrate distance, oxygen background pressure, heater temperature, laser fluence and laser spot areas [34]. In short, the stoichiometric transfer of ablated species to the substrate depends crucially on the plume dynamics. As demonstrated in several studies [35], some of them on the LAO/STO system [36], the overall stoichiometry of the films is strongly dependent on all parameters. For example, stoichiometric growth is preferred at high oxygen pressure since the kinetic energy of the cations is lower and the differences for light (Al) and heavy (La) species reduce. As a consequence, deviation from the ideal La/Al ratio in PLD grown LAO/STO films is common [37]. Moreover, a systematic molecular beam epitaxy study showed that conducting interfaces are created only in the case of La deficient samples. In particular, as shown in Fig.3, a finite conductivity at the $LaAlO_3/SrTiO_3$ interface is observed only for a La/Al ratio less than 0.97±0.03 [38].

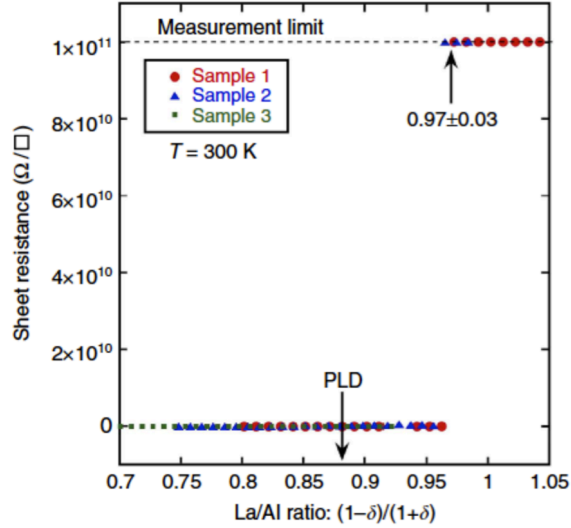

**Fig. 3.** Interfacial room temperature sheet resistance, obtained by local four-point probes as function of the La/Al ratio determined by RBS measurements of $La_{(1-\delta)}Al_{(1+\delta)}O_3$ films. A sharp jump in sheet resistance is observed at La/Al.0.97±0.03. An arrow indicates the stoichiometry of a PLD-grown companion sample. Reproduced with permission from ref. [38]. Copyright 2013 Nature Publishing Group.

Another crucial parameter determining the properties of LAO/STO system is the oxygen stoichiometry. The oxygen vacancies content is strongly dependent on the background oxygen pressure used during the deposition. Since the seminal work by Ohtomo and Hwang [6], it has been immediately recognized that films grown in $10^{-6}$ mbar or lower $O_2$ pressures, in particular when no post-oxidation process was employed, contain a substantial fraction of oxygen vacancies in the $SrTiO_3$ substrate, probably distributed in an inhomogeneous way throughout the whole crystal. These films show very low resistivity, which are essentially related to the important contribution to transport of carriers in bulk STO, also shown by the absence of the substantial anisotropy in the Shubnikov–de Haas oscillations of the magneto-resistance expected in the case of 2DEGs [39].

The oxygen vacancies content and the properties of LAO/STO interfaces depend on the oxygen pressure used during the deposition, as shown in Fig.4. Usually, with few exceptions [40], high oxygen pressures (> $10^{-3}$ mbar) produce insulating or barely conducting films, possibly related to a La/Al cation ratio close or larger than 1.

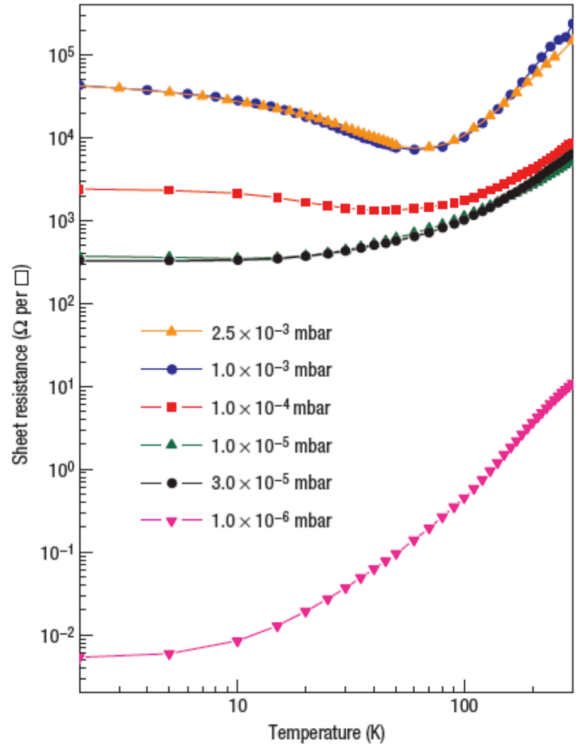

**Fig. 4.** Temperature dependence of the sheet resistance for n-type $SrTiO_3/LaAlO_3$ conducting interfaces, grown at various partial oxygen pressures (data from ref. [41]). Three regimes can be distinguished: low pressures leads to oxygen vacancies, samples grown at high pressures show a Kondo-like behaviour, whereas samples grown in the intermediate regime show superconductivity. Reproduced with permission from ref. [41]. Copyright 2008 Nature Publishing Group.

Films deposited between $10^{-3}$ mbar to $10^{-5}$ mbar are conducting and may contain small amount of oxygen vacancies. Post annealing in oxygen for films grown in this pressure range does not change substantially the transport properties, but slightly changes the carrier density and, as shown in the next paragraph, the density of localized electrons in $Ti^{3+}$ states. The latter influence the transport properties and can even behave as magnetic impurities, due to trapped electrons in localized states. The presence of magnetic $Ti^{3+}$ can lead to a Kondo-like transport at low temperatures or to the possible arising of magnetic effects, whose origin and nature are not completely understood. In particular, Brinkman et al. [41] reported low temperature (< 4 K) magneto-resistance hysteresis in films grown in $10^{-3}$ mbar and a Kondo-like upturn of the resistivity below 50 K. These results were ascribed to possible magnetic ground state of the system. On the other hand, samples de-posited in a 0.8-5 × $10^{-4}$ mbar range and post annealed in high oxygen pressure after the deposition are metallic and superconducting at low temperatures [8-10]. These LAO/STO interfaces are characterized by much smaller conductivity than non-annealed films, which is entirely due to the interface quasi-2D transport (no contribution from bulk STO). A small resistance upturn at low temperatures can be observed also in the case of these annealed LAO/STO heterostructures depending on the carrier density. However, as show in ref. [11], this behaviour is related to weak-localization that switches to anti-localization at higher doping due to Rashba spin-orbit scattering.

These results clearly suggest that oxygen vacancies and cation stoichiometry have an important role in the properties of this system. Both oxygen vacancies and La/Sr substitutions are indeed well known bulk STO dopant, but can behave also as scattering impurities. Intermixing or oxygen non-stoichiometry can be particularly likely at the LAO/STO interface and at the LAO surfaces. These defects can be partially eliminated by improving the growth process of LAO on STO and by systematically employing deposition conditions that allow good oxygen stoichiometric. Annealing after the deposition is a good method to reduce or eliminate oxygen vacancies, but similar properties can be achieved by using high oxygen pressures during the deposition and very slow cooling down after the growth in an oxidizing atmosphere. In any case, the presence of residual oxygen vacancies cannot be easily avoided. More complex is the control of the surface stoichiometry of $LaAlO_3$, which can have an important role as well (see next paragraph).

Moreover, the mobility of the q2DES in LAO/STO is strongly dependent on the deposition conditions and on the eventual presence of a capping layer. The record mobility values, up to $5x10^4$ $cm^2$ $V^{-1}$ $s^{-1}$ demonstrated by the Twente group, were obtained on LAO/STO bilayers capped by an epitaxial cuprate $SrCuO_2$ film [7].

Other STO interfaces exhibit q2DES with characteristics extremely similar to LAO/STO, namely: $LaGaO_3$/STO [42], $NdGaO_3$/STO [43], and $\gamma$-$Al_2O_3$/STO [44]. The latter is particularly interesting from a technological point of view since it shows a mobility exceeding $10^5$ $cm^2$ $V^{-1}$ $s^{-1}$.

## 3 Why a q2DES is formed at the LAO/STO interface ?

This paragraph is dedicated to a quick overview of the main mechanisms proposed to explain the formation of the q2DES at the interface between polar $LaAlO_3$ and $SrTiO_3$ band insulators. There are essentially three categories of possible theoretical approaches, those based on the so called "electronic reconstruction" scenario, which is an ideal picture explaining the formation of a q2DES in the case of perfectly stoichiometric LAO and STO layers; those based on extrinsic mechanisms, where the entire physics is attributed to extrinsic doping of the STO layers next to the interface, associated to possible cationic disorder and/or oxygen vacancies; finally, more recently, several theoretical approaches appeared in literature suggesting that the polar instability, which is the essential ingredient in the "electronic reconstruction" scenario, is the driving force for complex stabilization of the system that could take place by formation of specific defects at the interface and at the surface.

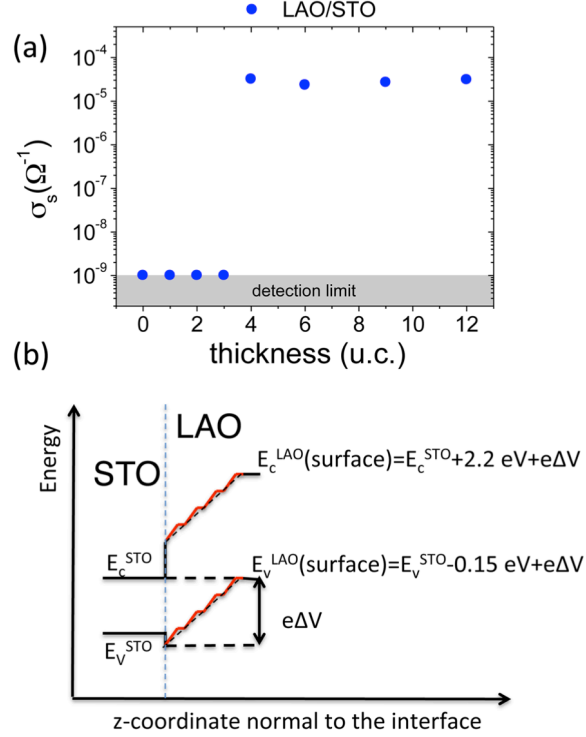

**Fig. 5. (a)** Sheet conductance of LaAlO$_3$/SrTiO$_3$ heterostructures as function of the LaAlO$_3$ thickness (in uc (u.c.)), grown by PLD at the CNR-SPIN labs. **(b)** A schematic band diagram of the LAO/STO interface just before the electronic reconstruction.

### 3.1 The electronic reconstruction mechanism

When LaAlO$_3$ (001) films are deposited on SrTiO$_3$ (001), in an ideal case, an electrostatic potential is accumulated. This is due to the dipole moment associated to charged LaO$^+$ and AlO$_2^-$ planes. This potential will rapidly grow as function of the number of LAO (001) unit cells, and will diverge in an infinite crystal giving-rise to a "polarization catastrophe". For perfectly flat planes, the LAO dipole moment amounts to $D_{bare}=-0.5e^- \times c_{LAO} = -1.9\ e^-$Å, which corresponds to an electrostatic potential $\Delta V=15$ V per each LAO unit cell deposited on STO. As matter of fact, the associated electrostatic energy would be larger than the LAO band gap of 5.6 eV already for one LAO layer deposited on STO. Thus an electrostatic instability is present already at the beginning of the growth. However, as pointed out by several papers, and in particular by Pentcheva and Pickett in ref. [45], the electric field can be partially screened by the ionic polarization of the LAO planes. Assuming a dielectric polarizability for LAO of $\varepsilon_r$ =24, the effective electrostatic potential energy will amount to about 0.20 eV/Å and it will equal the band gap only when the number of LAO layers is about 5 uc. Thus, according to simple electrostatic considerations a LAO (001) film can grow on STO (001) up to a critical thickness before some reconstruction has to take place. In agreement with this scenario, the q2DES appears only in samples characterized by a LAO thickness above 3 uc [10], as shown in Fig. 5a. This result, which was confirmed in many labs, is an important argument in favour of polar instability as main driving force for the formation of the q2DES at the LAO/STO interface. Now, it is generally agreed that any theoretical model should explain the formation of the q2DES at the critical thickness.

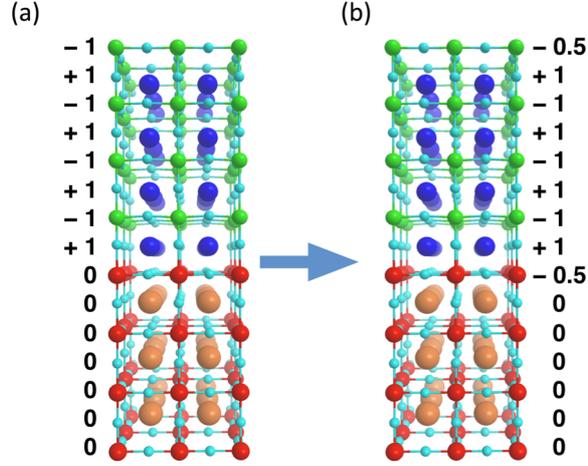

**Fig.6.** (a) Ideal stacking in the case of n-type LaAlO$_3$/SrTiO$_3$ interface, with indication of the formal charge of each planes (in unit of the elementary electron charge). (b) Formal charge of LaAlO$_3$/SrTiO$_3$ planes after electrostatic stabilisation which brings (formally) -1/2 charge from the AlO$_2$ surface to the TiO$_2$ interface.

One way to reduce the electrostatic potential at the critical thickness consists in charge balance at the LAO surface and at the LAO/STO interface, as shown in Fig. 6 in the case of an n-type interface, i.e. interfaces for which STO is TiO$_2$ terminated and LAO is ideally AlO$_2$ terminated (note that in this picture dangling bonds and (related) surface states are not considered). If the formal charge of the LAO surface and of the interface is reduced from -1 to -1/2 electron ($e^-$) per unit cell and, at the same time, the interface charge on the other side becomes -1/2 $e^-$, the electrostatic potential will not diverge anymore as function of the number of LAO layers.

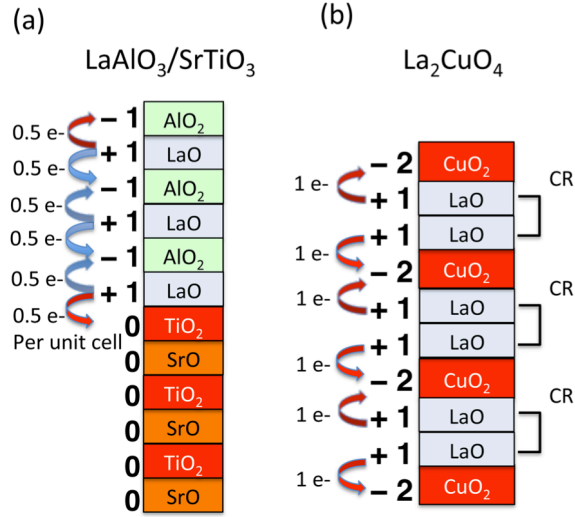

**Fig. 7.** (a) Stacking of $(LaO)^+$ and $(AlO_2)^-$ planes along the [001] direction of LAO/STO interface. Starting from neutral planes, at the equilibrium the LaO layers donate electrons which becomes bound in the neighbouring $AlO_2$ planes. In particular 0.5 electrons are donated to two neighbour $AlO_2$ planes in the bulk of LAO (on the upper and lower unit cells), and 0.5 electrons are donated naturally to the charge neutral $TiO_2$ plane (a positive uncompensated charge of -0.5 e should remain on the surface). (b) Similar electrostatic picture applied to the $La_2CuO_4$ cuprate. Each LaO layer transfer one electron to the $CuO_2$ planes so that Cu remains in a 2+ configuration, which corresponds to one hole in the 3d orbitals.

The redistribution of electrons from the LAO surface to the STO interface is believed favoured by the variable valence of transition metal titanium ions. Titanium in STO is in a $3d^0$ configuration and has a $Ti^{4+}$ valence. The transfer of $0.5e^-$ from the $AlO_2$ surface to Ti-3d states will change the valence of 50% of titanium ions at the interface from $Ti^{4+}$ to $Ti^{3+}$. DFT (density functional theory) calculations on n-type LAO/STO bilayers, taking into account the ionic polarizability of LAO, showed that the top of the valence band of the surface $AlO_2$ layers crosses the Fermi level at the critical thickness, thus at the same time electrons are transferred from LAO to the Ti-3d bands of the interface $SrTiO_3$ unit cells [45]. According to calculations this phenomenon should take place when exactly 5 uc of LAO are deposited on STO. The transfer of electrons at the interface stabilizes the system and reduces the electrostatic potential in the LAO film. Note that similar arguments hold in the case of p-type interfaces, i.e. in the case STO is SrO terminated and LAO consequently LaO terminated [46]. However, only n-type interfaces are conducting.

An alternative view of the polar instability scenario is illustrated in Fig. 7(a) [47]. In an ionic picture, where the equilibrium configuration of $LaAlO_3$ can be regarded as a stacking of $(LaO)^+$ and $(AlO_2)^-$ planes along the [001] direction, the LaO planes, starting from a neutral charge configuration, have to act as donor of electrons to the neighbouring layers. In particular 0.5 $e^-$ are donated and bound to neighbour $AlO_2$ planes in the bulk of LAO, so that ionic $(LaO)^+$ and $(AlO_2)^-$ planes are effectively stacked along the [001] direction, while at the interface 0.5 $e^-$ are donated naturally to the charge neutral $TiO_2$ planes (a positive uncompensated charge of -0.5 $e^-$ should in principle remain on the surface). This picture has the advantage of not requiring an unphysical mechanism, like Zener-tunnelling from the far away $AlO_2$ surface to the interface during the growth, to explain the appearance of compensating charges at both sides of the LAO thin film. Finally, this process is not at all uncommon in oxides. In cuprates [Fig.7(b)], for example, hole-doping of the $CuO_2$ planes is obtained by a charge transfer from the $CuO_2$ planes to the adjacent layers, which act as charge reservoirs. In the parent cuprate compound, like $La_2CuO_4$, starting from neutral LaO and $CuO_2$ planes, in the ionic equilibrium configuration a charge transfer from LaO to $CuO_2$ planes takes place. The charge-transfer process leaves one hole in the 3d orbitals per $Cu^{2+}$ site. By replacing a fraction $x$ of $La^{3+}$ by $Sr^{2+}$ will result in a number $x$ of missing electrons to be transferred to the planes (since SrO is neutral), leaving additional holes.

According to the ideal "electronic reconstruction scenario", the transition from insulating to conducting interfaces in LAO/STO is rather abrupt. It should manifest in a structural relaxation of LAO and STO layers at the critical thickness, and to an electric field in LAO varying with the thickness. Studies on the evolution of the electronic as well as of the structural properties of LAO/STO as function of the number of LAO layers have given important insights in the mechanism involved. Pauli et al. [48] and Cancellieri et al. [49] showed that qualitatively the evolution of the ionic displacements and of the c-axis of the LAO films agrees with the expectations based uniquely on the polar instability of the system. These results were confirmed in ref. [37], where it was shown that below 4 uc, i.e. in nominally 2 uc LAO/STO films, the LAO layers are strongly rumpled, indicating the presence of an internal electric field, and the unit cells are expanded in a direction perpendicular to the interface. However, while according to this scenario the STO interface layers should be essentially flat (no rumpling) before the critical thickness, we showed in ref. [37] the opposite behaviour, suggesting the presence of an electric field, localised charges, and consequent lattice distortions of the strongly polarisable STO layers before the appearance of a mobile electron system [Fig. 8].

Other important experimental results are in disagreement with the ideal electronic reconstruction scenario. Core-level photoemission spectroscopy did not detect a sizeable varying electric field with thickness [50]. Signatures of (nominal) $Ti^{3+}$ cations were found in insulating (2 uc) as well as conducting (thickness > 4 uc) LAO/STO samples [51]. Finally, the overall fraction of mobile electrons and of electrons in $Ti^{3+}$ states was found much lower than the value expected (0.5 e/u.c.) from the model, a puzzling result that is general referred in literature as the "missing charge problem".

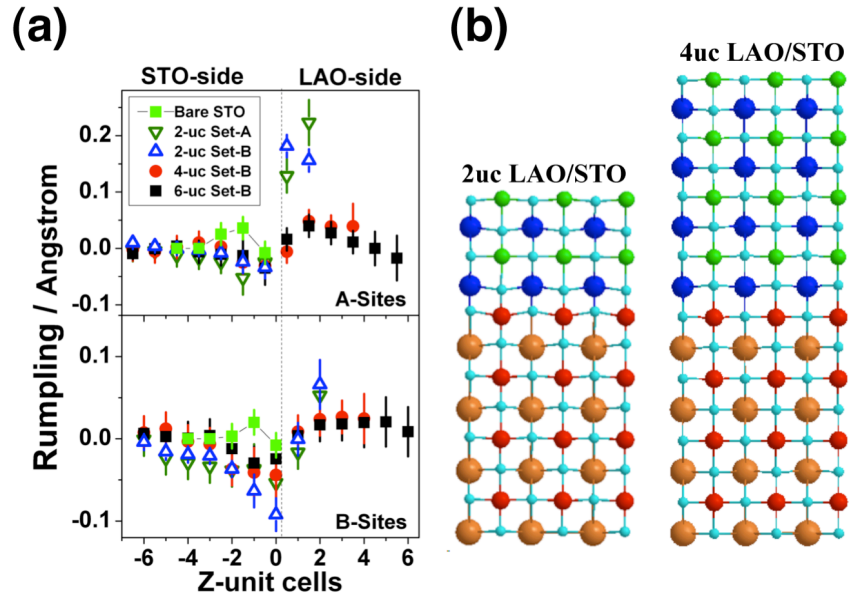

**Fig. 8.** (a) Results of the structural refinement of GXID data on LAO/STO samples, showing the rumpling of the AO (A-sites, upper panel) and $BO_2$ (B-sites, bottom panel) planes; closed green squares are for a bare STO surface, open symbols are for the 2-uc set-A (green, down triangles) and 2-uc set-B (blue, up triangles) samples, while closed symbols are for 4-uc (red circles) and 6-uc (black squares) set-B samples. (b) LAO/STO structures of 2-uc and 4-uc samples (set-B) as result of the structural refinement. Blue and green spheres are La and Al ions, while red and orange spheres are Ti and Sr ions. Oxygen ions are the small cyan spheres. Set A and Set B refer to samples grown at the University of Augsburg (group of Prof. J. Mannhart) and samples grown at the University of Geneva (group of Prof. J-Marc Triscone), respectively. Reproduced with permission from ref. [37]. Copyright © 2014 WILEY-VCH Verlag GmbH & Co. KGaA, Weinheim.

### 3.2 Other mechanisms

The strong argument in favour of the electronic reconstruction picture is the formation of the q2DES at a critical thickness, close to the predicted value. Thus alternative models should explain why the q2DES appears only at a critical thickness. This is quite important, since this feature clearly distinguishes the mechanism of formation of a conducting system in epitaxial LAO/STO from the recent reports of conductivity at amorphous/STO interfaces [52]. For amorphous/STO interfaces there is no clear critical thickness and the interfaces become insulating after annealing in $O_2$, so that for these systems oxygen vacancies are the most plausible explanation of the phenomenon [53].

On the other hand, an oxygen vacancies picture or a simple random introduction of La/Sr doping atoms at the interface (intermixing) is not able to ex-plain the existence of a critical thickness in the cases of epitaxial LAO films, deposited on $TiO_2$ terminated STO, for two reasons: first, the density of defects are expected strongly dependent on the deposition conditions, so should be the critical thickness that is not the case; both oxygen vacancies and intermixing cannot change the interfacial charge balance, since both of them are globally neutral defects. In particular, for each La(Sr) or Al(Ti) exchange, the interface will get both donor and acceptor levels (thus electron and holes) which are expected to be annihilated being located essentially at the same place. Note that cation intermixing were observed by high-resolution transmission electron microscopy (HRTEM) [54] and by grazing incidence surface diffraction (GXID) by several groups on samples grown in different laboratories [37, 55]. As matter of fact the degree of La/Sr and Al/Ti substitution was found dependent on the growth conditions, the sample thickness, and varied from sample to sample. There is at the moment no systematic study on the relationship between conductivity and cation disorder (at fixed composition) in this system, but it would be strongly desirable. The general idea is that intermixing alone cannot explain the critical thickness issue (it should be accompanied by off stoichiometry). Similar arguments hold for oxygen vacancies in the STO layers, which can be present even in oxygen annealed LAO/STO interfaces to some degree (see ref. [37] and Fig. 9), but it does not compensate the charge polarity, thus leaving the polar instability problem unsolved.

In order to explain both the critical thickness result and the number of experiments disagreeing with the ideal "electronic reconstruction" scenario, recently other approaches have been proposed, which take into account the polar instability and the energy necessary to form in these conditions various kind of non-stoichiometric defects. A completely overview of these approaches can be found in two recent papers, which discuss the problem of formation of a q2DES at the LAO/STO interface in quite complete fashion, but with different approaches [56, 57]. Here we just argue that the most likely defects, which can effectively at the same time give a solution of the polar instability problem and can explain the formation of a conducting LAO/STO interface, are oxygen vacancies at the LAO surface, also considered in recent papers [58]. The formation energy is, according to theoretical calculations, dependent on the electric field accumulated in LAO, so that only at a critical thickness, comparable to the one experimentally observed, such defects are stable in oxidising atmosphere at the $AlO_2$ surface. This idea has the advantage of explaining, for example, a certain sensitivity of the q2DES to surface treatments (including atomic force microscopy tip writing [59]). However a definitive assessment of this picture has to be validated. For example, by using GXID in ref. [37] we have found that the LAO surface is always incomplete, independently on the thickness, while allowing oxygen vacancies as possible refinement parameter, the amount of these defects are found not much dependent on the thickness and present in all the LAO layers at level of less than 2% down to the first interfacial unit cell (note that the sensitivity of x-ray diffraction to light elements like oxygen, is limited and error bar in the estimation of oxygen vacancies is quite large). Systematic in-situ studies of the conductivity and surface/chemical properties of LAO/STO after different surface treatments are missing, but probably they could provide crucial insight in the physics of this system.

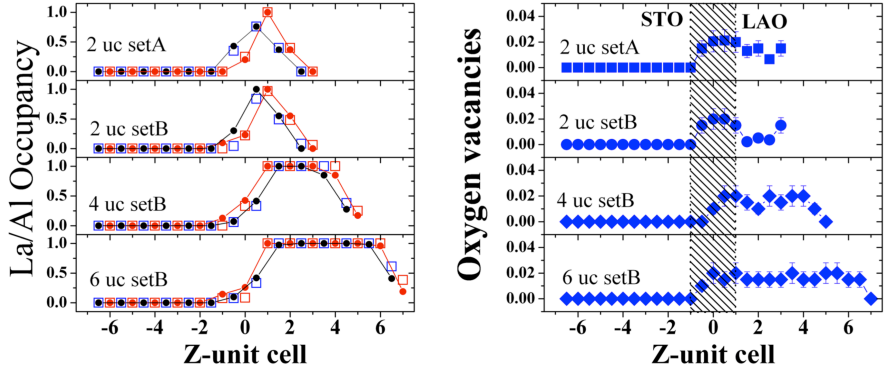

**Fig. 9.** (a) La/Al occupancies of the LAO/STO bilayers from structural refinement using two structural models: model-1, full oxygen occupancy (closed circles); model-2, oxygen vacancies model (open squares). Red (site-B) and black (site-A) closed circles are the result from model-1 while red (site-B) and blue (site-A) open squares are the occupancies obtained using model-2. (b) Corresponding oxygen vacancy level from model-2 for each sample as function of the layer-index. Dashed regions correspond to areas where cation intermixing is found from the refinement. Set A and Set B refer to samples grown at the University of Augsburg (group of Prof. J. Mannhart) and samples grown at the University of Geneva (group of Prof. J-Marc Triscone), respectively [37].

## 4. Spectroscopy studies of the electronic properties of LaAlO$_3$/SrTiO$_3$ heterostructures

Spectroscopy techniques are fundamental tools for the investigation and understanding of condensed matter physics problems. In particular the use of synchrotron radiation and scanning probe microscopy/spectroscopy are having an important role in clarifying the ground state of the q2DES at the LAO/STO interface and the mechanism leading to its formation.

One of the main difficulties in the investigation of this system by using spectroscopy probes, is the fact that the region of interest is buried, thus not directly accessible. This section will present some of the results obtained by spectroscopy techniques on the LAO/STO system and will analyse these results in view of the proposed theoretical mechanisms discussed in Section 3.

### 4.1 X-ray Absorption and X-ray Photoemission spectroscopies

X-ray Absorption Spectroscopy (XAS) has been used to study the electronic, magnetic properties and the crystal-field splitting of LAO/STO films across the insulating to metal transition. XAS is a probe of unoccupied density of states of a given ion in a crystal structure. XAS spectra depend on many factors, including crystal field splitting, valence oxidation state of the absorbing ion, and subtle effects like the effective spin-orbit coupling in the final state and charge transfer effects which can partially account for the hybridization between oxygen 2p and transition metal 3d-states. Fig.10a shows a schematic of the XAS process in the case of x-rays with energies resonant with Ti-$L_{2,3}$ absorption edges. A 2p electron absorbs a photon with a given polarization. The 2p electron is excited into empty Ti-3d states leaving a core-hole. The XAS spectra depend on the density of unoccupied states and on the polarization. Using total electron yield method of detection, this technique is able to explore the interfacial states, with a probing depth of the order of 4 nm.

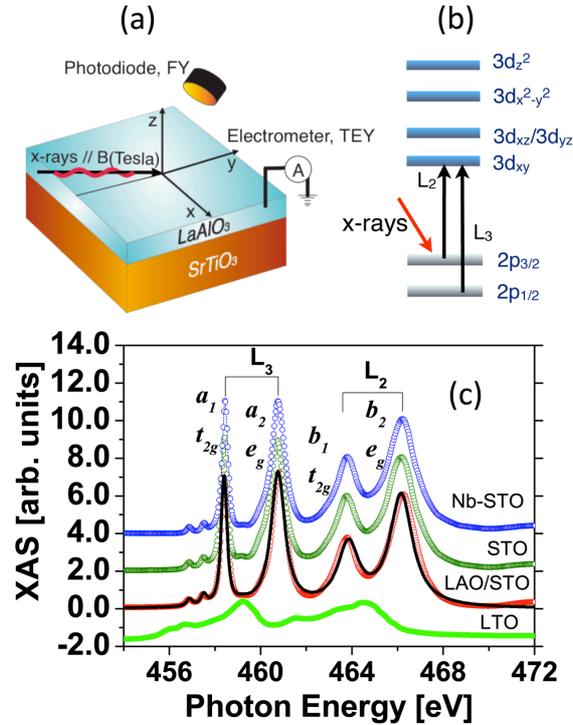

**Fig. 10.** a) Sketch of the XAS process in the case of LAO/STO interfaces for $L_{2,3}$ Ti-edge. b) XAS spectra measured at the ID08 beam-line of the European Synchrotron Radiation Facility (ESRF): data for LTO (green circles) conducting LAO/STO (red circles), insulating STO (dark green circles) and conducting 1%at Nb-STO (blue circles) are displaced by fixed amount and normalized at the $L_3$ absorption edge; atomic multiplet calculations reproducing perfectly the data assuming a pure $Ti^{4+}$ oxidation state are shown as black continuous line.

According to the electronic reconstruction scenario (Section 3.1), 50% of interfacial titanium ions are expected to change valence from $Ti^{4+}$ to $Ti^{3+}$ due to a transfer of $0.5e^-$ from the $AlO_2$ surface to Ti-3d states. As shown in Fig.10b, oppo-site to the expectations, the XAS spectra of LAO/STO samples, of insulating and of conducting (Nb-doped) SrTiO$_3$ single crystals are very similar to each other and, as matter of fact, fully reproduced by atomic multiplet splitting calculations using crystal field parameters typical of insulating STO, i.e. 3d$^0$ ground state configuration for titanium ions in Ti$^{4+}$ oxidation state. The complex shape of Ti-edge XAS spectra, characterized by four main peaks (***a$_1$, b$_1$, a$_2$, b$_2$***), is related to the fact that, in a 3d$^0$ configuration, final t$_{2g}$ and e$_g$ states are empty. On the contrary, the antiferromagnetic Mott insulating LaTiO$_3$, which is the closest realization of 3d$^1$ Ti$^{3+}$ with one electron per site localized in 3d-Ti states, is characterized by XAS spectra with much broader L$_{2,3}$ peaks located at energy values in the middle between ***a$_1$, b$_1$*** and ***a$_2$, b$_2$***.

These results apparently contradict the conducting character of the interface, which implies the presence of delocalized electrons in the 3d bands. Moreover, core-level hard x-ray photoemission (HXPES) experiments have given evidence of the presence of a satellite Ti2p shoulder [51], as shown in Fig. 11a, which is commonly attributed to the emission from Ti ions in a 3+ oxidation state. However, XAS and HXPES data can be easily reconciled considering the exact nature and differences in XAS and core level XPS processes (for a complete overview about the differences between XAS and XPS core level spectroscopies see ref. [60]). In both cases, a 2p electron is excited into unoccupied 3d-states, leaving a core-hole, which modifies substantially the local electronic configuration of the absorbing ion through the appearance of a core-hole potential U$_{cd}$. However, the effect of the core-hole potential is different in XAS and XPS. In XPS, the core-hole interaction gives rise to a satellite in the XPS spectra due to the effect of U$_{cd}$ on the energy of emitted 3d electrons. In XAS, the core-hole interaction is counterbalanced by the Coulomb repulsion, U$_{dd}$, between excited and 3d electrons (unless the ground state is 3d$^9$ as in cuprates). Consequently, with the restriction U$_{dd}$=U$_{cd}$, the satellite spectral weight is essentially null, thus electrons in charge transfer states do not give rise to a satellite. In the general case (U$_{cd}$≠U$_{dd}$), the transition probability is proportional to the U$_{dd}$-U$_{cd}$ difference, thus XAS is in any case less sensitive to itinerant electrons than XPS. Another way to visualize the differences between XAS and XPS is the following: in the XPS process electrons in the Ti-2p bands, excited to 3d states, can be screened by carriers in the 3d bands. Coherently screened electrons have energy different from unscreened ones, and thus give rise to a satellite peak in the spectra. In the XAS case, the core-hole excitation is globally neutral, thus excited electrons into 3d states are not (less) screened. Thus, delocalized electrons in Ti-bands hardly show any signature in x-ray absorption spectroscopy, unless they are in such a density to change substantially the effective oxidation state of a large fraction of titanium ions (>10%) from 4+ to 3+.

On the contrary, XAS is able to detect localized electrons in Ti$^{3+}$ states, as in the case of LaTiO$_3$ and LAO/STO interfaces containing a substantial fraction of oxygen vacancies. The latter are artificially introduced by growing LAO/STO samples in standard conditions (typically using P$_{O2}$=10$^{-4}$ mbar of oxygen pressure and deposition temperatures of 780°C) without high oxygen pressure post-annealing [61]. In these cases, we can notice that (Fig. 11b) the XAS spectra show a transfer of spectral weight from $a_1$, $b_1$ (and from $a_2$, $b_2$) features, typical of a $Ti^{4+}$ valence, to the middle regions where contribution from $Ti^{3+}$ is expected. This result is in good agreement with XPS data of Fig. 11a [51], showing that non-annealed samples, which contain oxygen vacancies, are characterized by a substantial increase in the satellite Ti-2p core level peak.

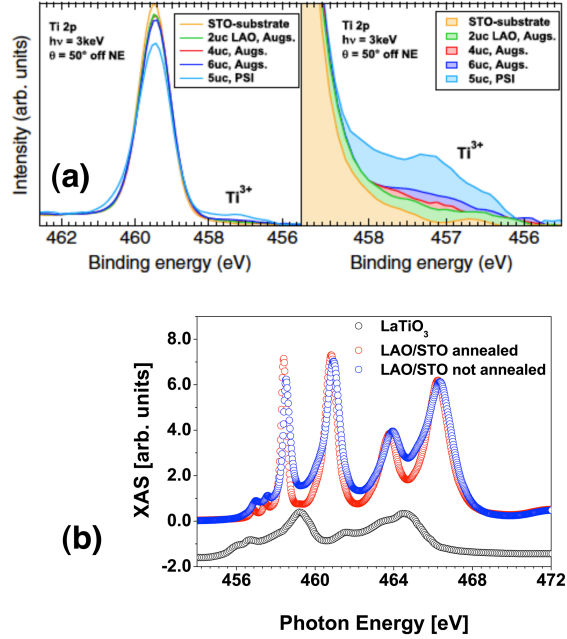

**Fig. 11.**: (a) Ti 2p HXPES spectra of various samples plus bare STO at a fixed emission angle. PSI LAO/STO samples are grown in a reducing atmosphere without annealing in oxygen after the deposition. Augsburg (Augs) samples have been grown by PLD in $10^{-5}$ mbar of $O_2$ partial pressure deposition conditions and annealed in oxygen after the deposition [Reproduced with permission from ref. [51]. Copyright © 2009 American Physical Society]. (b) XAS data on LAO/STO samples annealed (red circles) and non-annealed (blue circles) in oxygen after the deposition (from Ref. [61]). Data for LTO are also shown as reference.

**4.2 Linear Polarization dependent X-ray absorption: the Orbital reconstruction**

Using linearly polarized photons in combination with geometrical configuration in photon-in electron-out (or photon-out) spectroscopy experiments one can have information on the anisotropy of the Ti-3d states and on the splitting of the bands. In particular, the x-ray linear dichroism (XLD), i.e. the difference between XAS spectra acquired with a polarization perpendicular ($I_c$) and parallel ($I_{ab}$) to the interface, is a direct measure of the orbital polarization and provides, through direct comparison with calculations, the splitting between in-plane and out-of-plane $t_{2g}$ ($3d_{xy}$, $3d_{xz,yz}$) and $e_g$ ($3d_{x2-y2}$, $3d_{z2}$) orbitals. The XLD spectra of STO and LAO/STO interface, shown in Fig.12a, demonstrate a very intriguing inversion of the energy levels. While the XLD spectra of a $TiO_2$ terminated STO single crystal exhibit crystal field parameters equal to $\Delta t_{2g}=3d_{xz,yz}-3d_{xy}=+25$ meV and $\Delta e_{2g}=3d_{z2}-3d_{x2-y2}=+40$ meV (splitting of unoccupied $Ti^{4+}$ 3d states), the sign of the dichroism changes in LAO/STO interfaces. In particular when the LAO thickness becomes equal to 4uc and the interface becomes conducting, the amplitude of the XLD signal saturates and the corresponding 3d splitting is $\Delta t_{2g}=3d_{xz,yz}-3d_{xy}=-50$ meV and $\Delta e_{2g}=3d_{z2}-3d_{x2-y2}=-90$ meV [62]. It is important to note that the data are perfectly reproduced by atomic multiplet splitting calculations assuming a $Ti^{4+}$ ionic configuration, and not $Ti^{3+}$ as expected in the electronic reconstruction scenario.

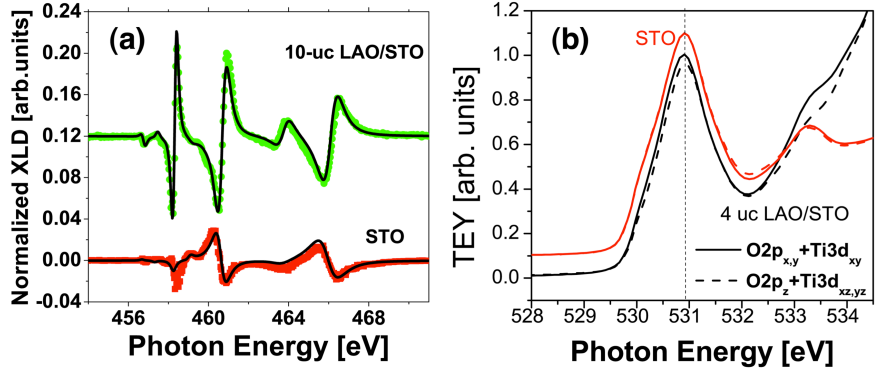

*Fig. 12.* (a) Ti $L_{23}$ edge XLD (Ic-Iab) spectra of STO (red symbols) and conducting LAO/STO (green symbols) (from ref. [62]). Black lines are atomic multiplet calculations assuming a $Ti^{4+}$ configuration. (b) O-K edge XAS spectra in TEY mode for insulating STO and conducting LAO/STO samples. Continuous lines are acquired with polarization parallel to the ab plane. Dashed lines are acquired with polarization perpendicular to the interface.

The removal of the orbital degeneracy is also reflected in the relative contributions of in-plane and out-of-plane orbitals to the conduction band of the interface. A first direct indication of band splitting comes from oxygen K-edge XAS spectroscopy, as shown in Fig. 12b. Since the oxygen 2p states are nominally filled in both LAO and STO, a 1s→2p transition occurs only because the oxygen 2p states are hybridized with ligand cation orbitals in the system, including 3d Ti-orbitals. Because of this hybridization, the peak A of Fig. 12b at 531eV is related to the interface conduction band, located in STO just above the Fermi level, and composed of Ti-3d $t_{2g}$ states hybridized with oxygen 2p states [63]. While no polarization dependence is observed for an STO single crystal, the absorption line of conducting LAO/STO samples shifts by about 50 meV toward the Fermi level when the photon polarization is parallel to the xy-plane [64]. In the xy-polarized spectra the peak A represents the contribution to the conduction band of in-plane titanium $3d_{xy}$ hybridized with oxygen $2p_{x,y}$ states. Its shift towards the Fermi energy shows that the $3d_{xy}$-O2p band is the lowest in energy. Similar conclusions were obtained by polarization dependent resonant angle resolved soft x-ray photoemission results shown in ref. [65], where the authors found that the bottom of the occupied $3d_{xy}$ band at the Γ-point is lower than those of the $3d_{xz,yz}$ bands, in analogy to similar data reported on STO (001) surface, also showing a 2DES [66, 67].

These experimental data show that an orbital reconstruction, characterized by an inversion of the bands respect to the bulk, is a distinctive feature of the LAO/STO interface. However, the origin of the orbital reconstruction is not yet completely assessed. According to calculations [68, 69], the interface-induced asymmetry and the quasi-2D-nature of $3d_{xy}$ orbitals decrease the energy of the $3d_{xy}$ band compared to the out of plane $3d_{xz,yz}$ bands. However, the inversion of energy levels takes place already at a LAO thickness below the critical value, when the interface is still insulating [37]. It is interesting that the orbital reconstruction is accompanied by the development of a substantial rumpling of the $TiO_2$ planes, which have all the characteristics of a polarization response of STO to the presence of an electric field [37]. Another key feature of LAO/STO heterostructures is the presence of a large spin-orbit Rashba splitting [11]. It is very likely that the orbital reconstruction, the rumpling of the $TiO_2$ planes and enhanced spin-orbit interaction are closely related.

These experimental findings have important consequences on the ground state of the system. In particular, in LAO/STO it is possible to achieve regions of doping where electrons occupy uniquely $3d_{xy}$ orbitals (which cannot be achieved in bulk STO). One can then envisage multi-band orbital effects modifying the characteristics of the system, and conferring unique properties to the q2DES among other 2D-systems based on semiconducting, non-oxide, materials.

**4.3 Scanning Probe Microscopy/Spectroscopy**

The main difficulty in investigating the electronic properties of the LAO/STO interface is the fact that the interface is buried below at least 4 uc, i.e. 1.5 nm, of LAO. However, LAO is insulating, thus tunnelling experiments are able to get information on the local density of states even using standard planar geometries, where the tip is above the surface. The exponential attenuation of the tunnelling current by the LAO barrier requires the use of reduced tunnelling currents and, to avoid contribution from the LAO surface, of bias voltages well below the conduction band minimum (CBM) of LAO (i.e. $V_{bias} < 2.2$ V).

The first attempts to measure the density of states of the LAO/STO interface by Scanning Tunnelling Microscopy, were performed by M. Breitschaft et al. [70] at cryogenic temperatures. Through comparison between experimentally obtained density of states and theoretical modelling, M. Breitschaft et al. proposed an interesting model for the q2DES, suggested to be a 2D-electron liquid due to quantum confinement perpendicular to the interface. The possible role of electron correlations have been highlighted also by other experiments, and in particular by recent studies of the superconducting order parameter using planar tunnelling spectroscopy [71].

A detailed study of spatial and interface selective density of states at room temperature was performed by using an ultra high vacuum scanning probe spectroscopy set-up, operated in STM mode. STM data were collected on LAO/STO films characterized by a thickness of 4 uc grown at the University of Augsburg in standard conditions and annealed in oxygen after the deposition. Tunnelling spectroscopy data were acquired by using ultra-low currents in the sub-picoampere regime [64, 72].

Two main results were obtained from spatially resolved local density of states (LDOS) data. First, at positives bias values (in the 0 to +1.5 eV range) the interface shows a quasi-periodic corrugation, which corresponds to a modulation of the unoccupied LDOS [Fig.13] and in particular to a different local shift of the 3d conduction band toward the Fermi level. In this energy region, the contribution to the LDOS is due to unoccupied 3d-$t_{2g}$ Ti states. The main features observed are two characteristic changes of the slope at +0.5 eV and +1.2 eV, which nicely match calculated partial DOS from $t_{2g}$ 3d orbitals (see for example ref. [68]). In the occupied region of the spectra, the data show in a range from 0 to -1.5 eV a LDOS which increases in a continuous way, thus showing not only the presence of states at the Fermi level, but also states in the gap. In particular, locally the LDOS is characterized by a peak at -1.3 eV, which is intriguingly similar to in-gap states observed by photoemission spectroscopy at the STO surface [66] and later-on at the LAO/STO interface [58]. A statistical analysis of the occurrence of this in-gap features at the interface, reveals that around 40% of the interface is characterized by electrons localized in the gap [72].

It is worth noting that oxygen vacancies cannot explain the large fraction of the interface showing an in-gap feature, unless one would assume an unphysical large fraction of these defects in oxygen annealed LAO/STO samples (in contradiction with other experiments). Thus, the in-gap states have been interpreted as an intrinsic feature of the LAO/STO system, and in particular of the process of electron doping at the titanate surfaces/interfaces: instead of transferring all the electrons into coherent 3d bands at the Fermi level, a large fraction of these electrons are trapped in states in the gap to compensate the polar discontinuity at the interface. Assuming one electron per site, the fraction of 0.4 $e^-$/uc counted by STS spectroscopy is not far from the 0.5 $e^-$/uc predicted by the electrostatic instability scenario. Moreover, electrons in these in-gap states have energy quite far from the Fermi level and from the 3d-Ti conduction bands, thus they have probably mixed Ti3d and O2p character, as it happens at the STO surface [73]; this interpretation is consistent with the fact that the fraction of observed $Ti^{3+}$ is much lower than the value expected, since electrons are not localized in $Ti^{3+}$ states, but in states which are obtained from the hybridization between O2p and Ti-3d orbitals.

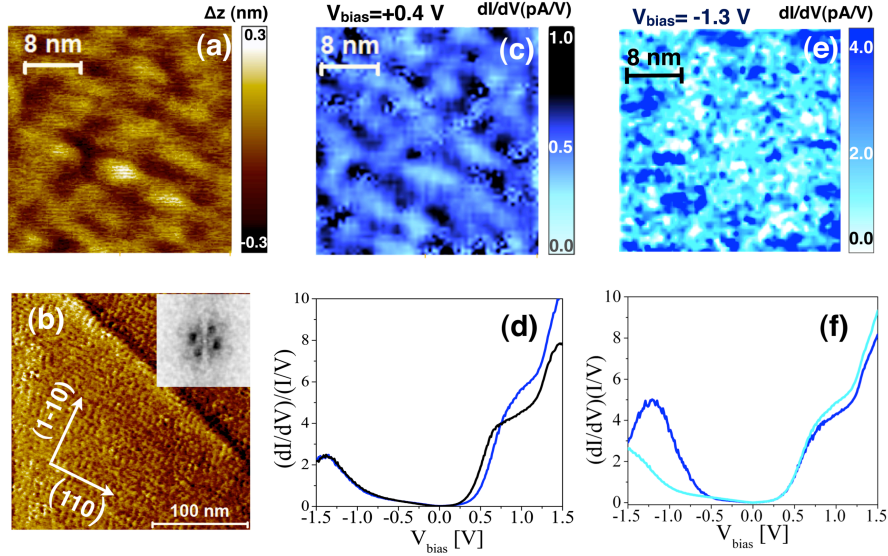

**Fig. 13.** STM/STS data on LAO(4uc)/STO conducting interface [64]. STM topography on (a) a 32x32 nm$^2$ and (b) on a 256x256 nm$^2$ area ($V_{bias}$=+0.9 V, $I_t$=0.75 pA). Inset of (b) shows power fast fourier transform of (b). dI/dV map taken simultaneously with topography in (a) at (c) +0.4 V and (e) -1.3 V. In (d) normalized (dI/dV)/(I/V) spectra averaged over two classes defined by low (blue) and high (black) tunneling conductance in (c). (f) Normalized (dI/dV)/(I/V) spectra averaged over two classes defined by a peak (cyan) and normal (cyan) tunneling conductance at -1.3 V from the map in (e).

### 4.4 Magnetism at the LaAlO$_3$/SrTiO$_3$ interface

The ground state of the LaAlO$_3$/SrTiO$_3$ system is one of the most interesting but still controversial issue due to apparently conflicting observations of superconductivity of the q2DES below 0.3 K in some samples [8] and of magnetic effects in others [41]. The stabilization and control of an interface magnetism based on titanate interfaces would be, on the other hand, a major technological achievement since it could provide an oxide platform for spin-polarized 2DEG. According to early theoretical studies, titanate heterostructures can become ferromagnetic and metallic by an interface orbital/magnetic reconstruction. There are two crucial conditions which can led to the effective realization of a ferromagnetically ordered state: the electrons transferred to the otherwise empty titanium 3d-sites modify the titanium nominal valence from Ti$^{4+}$ to Ti$^{3+}$ and the orbital occupation of 3d states from 3d$^0$ to 3d$^1$; the 3d orbital degeneracy should be removed in such a way that 3d$_{xy}$ and 3d$_{x2-y2}$ states are pushed below the out of plane 3d$_{xz,yz}$ and $3d_{z^2}$ orbitals. The LAO/STO interface undergoes the orbital reconstruction predicted by theory, so that electrons transferred to the interface preferentially occupy $3d_{xy}$ states. The idea is that these electrons, on the other hand, carrying not only a charge but also a spin, in the presence of electron correlations, can order ferromagnetically. Recently, SQUID (Superconducting Quantum Interference device) [74], torque magnetometry [75] and scanning SQUID microscopy [76] reported evidence of coexistence of magnetism and superconductivity at low temperatures. However, these techniques cannot provide the direct proof that magnetism is indeed an intrinsic phenomenon related to $Ti^{3+}$ moments at the interface, because they are not able to distinguish the magnetic signal coming from the interface compared to inner and topmost (LAO*)* layers, and are not elemental and orbital selective. As matter of fact, the evidences of robust magnetism in this system conflict with other experimental reports; among them polarized neutron reflectometry [77] and β-detected nuclear magnetic resonance [78] have shown that any magnetic moment in this system is very small, not compatible with the large values (up to 0.3 $\mu_B$/Ti at the interface) estimated by torque magnetometry experiments and by scanning squid microscopy.

X-ray magnetic circular dichroism at the Ti $L_{2,3}$ edge is probably one of the few experimental techniques able to provide the proof of a purely titanium interfacial magnetism, having at the same time the characteristics to provide both elemental, orbital and interface selective information. Moreover, it is one of the most sensitive magnetic techniques since it is able to detect the magnetic moment of transition metal adsorbates with concentration below 1% [79].

In Fig.14 we show simulated XMCD spectra of $Ti^{3+}$ and $Ti^{4+}$ ionic configurations, calculated using atomic multiplet splitting models in C4 symmetry and a magnetic exchange of +10 meV. The position of the main features in simulated XMCD spectra have some correspondence with the maximum of intensity in the corresponding XAS spectra, and are quite different in the case of $Ti^{4+}$ or $Ti^{3+}$ oxidation states. In the case of a $Ti^{4+}$ configuration, a finite XMCD signal is related to the splitting of spin-up and spin-down unoccupied states in the presence of an exchange interaction. A purely $Ti^{4+}$ XMCD spectra does not relate to an effective magnetic moment, but just to an asymmetry of the spin-up/spin-down 3d density of states. On the other hand, a $Ti^{3+}$ XMCD signal different from zero is related to the presence of an effective magnetic moment of $Ti^{3+}$ electrons.

Lee et al. have performed XMCD experiments on 3 uc (and 10 uc) LAO/STO sample and found, in magnetic field of 0.2 Tesla parallel to the interface, a magnetic signal attributed to $Ti^{3+}$ $3d_{xy}$ electrons in the interfacial layer [80]. However, the data of ref. [80] show main features at energies corresponding to the peaks of the XAS spectrum at *$a_1$*, *$b_1$* and *$a_2$*, *$b_2$*, typical of a $Ti^{4+}$ oxidation state. The XMCD spectrum of ref [79] can be reproduced assuming the presence of $Ti^{3+}$ magnetic moments and charge transfer effects [Fig. 14, red line bottom panel] after a rigid energy shift of the calculated spectrum.

A different interpretation of the titanium XMCD data arises from a systematic work performed on LAO/STO samples in both strongly and poorly oxidizing conditions [61]. Standard LAO/STO samples, grown in $P_{O2}$=8 x$10^{-5}$ mbar and annealed in 200 mbar of $O_2$ after the deposition were compared to non-standard LAO/STO, which were not annealed after the deposition, in order to introduce some amount of oxygen vacancies at the interface and related localized $Ti^{3+}$ electrons. The experimental results were crosschecked by carrying out experiments at two x-ray synchrotron facilities, and in particular at the beam-line ID08 of the European Synchrotron Radiation Facility and at the X-TREME beam-line of the Swiss Light Source.

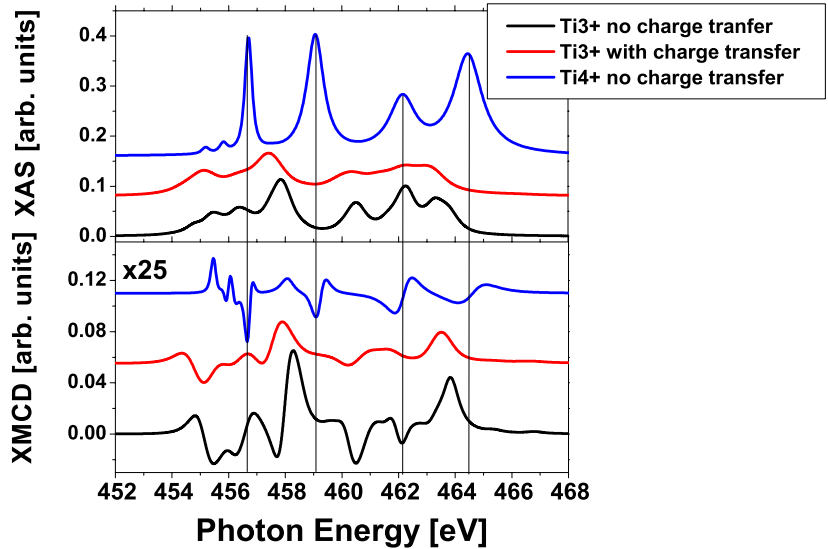

**Fig. 14.** Simulated XAS (upper panel) and XMCD (bottom panel) spectra of $L_{23}$ Ti-edge in $Ti^{3+}$ and $Ti^{4+}$ configurations using the atomic multiplet scattering code CTM4XAS [81]. Blue lines are obtained assuming a $Ti^{4+}$ oxidation state of titanium, red and black data are obtained assuming a $Ti^{3+}$ (with and without charge transfer respectively) oxidation state. The calculations have been performed in C4 symmetry (equivalent to octahedral configuration) and an effective exchange field of +10 meV. The charge transfer parameters, 10Dq and slater integral are taken from Ref. [61] and Ref. [80] in the case of $Ti^{4+}$ and $Ti^{3+}$ configurations respectively.

In Fig. 15a we show a summary of the experimental results.

The magnetic (both orbital and spin) moment is found negligible in the case of optimally oxygenated LAO/STO interfaces. On the other hand, samples containing oxygen vacancies show a signal related to $Ti^{3+}$ localized spins, as seen by features resonating with energies typical of $t_{2g}$ $Ti^{3+}$ XAS peaks. Thus oxygen vacancies introduce localized and also magnetic $Ti^{3+}$ 3d-electrons. Together with a $Ti^{3+}$ signal, these non-annealed LAO/STO samples are characterized by XMCD with features resonating also at the main $Ti^{4+}$ peaks suggesting a spin-splitting of the $3d^0$ states: excited electrons with up and down spins, giving rise to a finite XMCD signal.

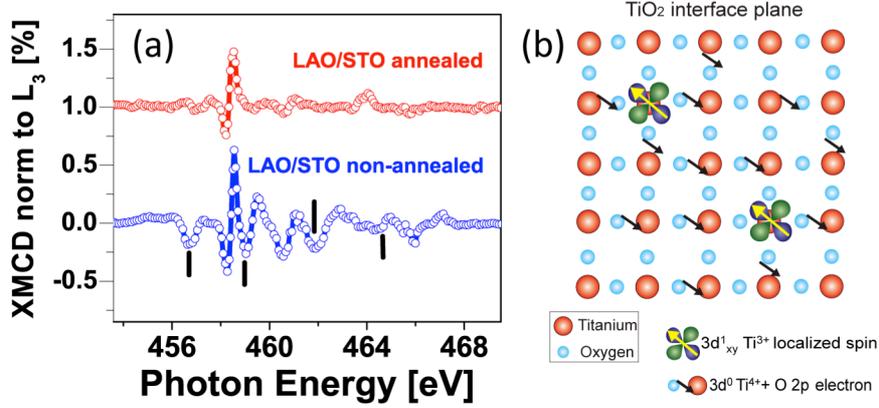

**Fig. 15. :** (a) XMCD data measured at 3K and 6 T on annealed (red symbols) and non-annealed( blue symbols) LAO/STO interfaces. Black sticks indicate features associated to localized $Ti^{3+}$ magnetic moments. (b) A cartoon illustration of the proposed mechanism of magnetism at the titanate heterostructures. In LAO/STO oxygen vacancies lead to localized $Ti^{3+}$ sites, holding sizable spin and orbital magnetic moments thanks to their $3d_{xy}$ electron. The q2DES gets polarized by these magnetic impurities and can mediate a long-range magnetic interaction among them at low temperatures.

Very strikingly, according to atomic multiplet splitting calculations, the sign of the features resonating with $Ti^{4+}$ XAS peaks is opposite to what is expected in the case of a positive exchange (e.g. for trivial Zeeman splitting, Fig. 14a blue line), and as matter of fact qualitatively reproduced by a negative magnetic exchange [61]. This result suggests that the titanium 3d-bands are spin-polarized through a negative exchange interaction between 3d electrons, including those forming the q2DES, and localized $Ti^{3+}$ magnetic moments, as shown schematically in Fig.15b. Since a purely $Ti^{3+}$ magnetic moment is detected only in the case of samples containing oxygen vacancies, it is possible to rule out ferromagnetism as an intrinsic property of this system.

     A tiny ($Ti^{4+}$) XMCD signal is observed also in the case of LAO/STO samples annealed in oxygen after the deposition, thus containing only small amount (if any) of oxygen vacancies. Here, the Ti-XMCD and XAS spectra cannot be explained by the presence of $Ti^{3+}$ magnetic moments. Yet, at $L_3$ and for $t_{2g}$ states, the XMCD is different from zero even at 0.1 Tesla. Any criteria used to get some quantitative information on the magnetic moment tell that the effective spin and orbital moment are extremely small (<0.001 $\mu_B$/Ti). Thus, the XMCD data in the case of annealed LAO/STO interface could be just related to an asymmetry between spin-up and spin-down density of states, i.e. a spin polarization of the q2DES, but not necessarily to a magnetic order. This is an intriguing result, which requires further investigations, also in view of recent theoretical proposal of spiral magnetism in q2DES [82].

## 5. Discussion and concluding remarks

This chapter was dedicated to a review of the general properties and on the open issues on the research on q2DES at the oxide LAO/STO interfaces. The main mechanisms proposed to explain the phenomenon were discussed and the main discrepancies, as well as agreements, with experiments were highlighted.

The main emphasis of this review was on the comparison between spectroscopy investigations of the LAO/STO system, the mechanisms proposed to explain the phenomenon, and the ground state properties. In particular, some of the weaknesses and strengths of the ideal "electronic reconstruction" scenario, largely based on the polar instability of polar/non-polar oxide interfaces, have been discussed.

The importance of the polar-discontinuity should not be questioned. Thus, starting from the ideal system, an interfacial charge of 0.5 $e^-$ should be transferred from LAO to STO in order to compensate this discontinuity. In some sense we can say that this interfacial charge is topologically protected by electrostatics. However, how this compensation takes place is in question. The roles of La/Al stoichiometry, of cation as well as oxygen vacancies at the interface and at the LAO surface are certainly very important, and possibly crucial to the improvement of mobility in this system.

The mobile charge measured by Hall effect is much lower than the value required to solve the polar catastrophe. The valence change of titanium from $Ti^{4+}$ to $Ti^{3+}$ is only partial and, in the case of samples annealed in oxygen, much reduced. XAS results have shown that the fraction of localized electrons in $Ti^{3+}$ $3d^1$ states is quite small, much lower than the 0.5 $e^-$/Ti expected. Similarly, the position of core-level XPS peaks associated to La and Al in LAO/STO as function of the thickness, a measure of the internal LAO electric field, is very small compared to results predicted by theory. The "missing charge" problem, highlighted in section 3, is certainly one of the most important, not-understood, issues in an ideal "electronic reconstruction" scenario. A trivial defect-related explanation of trapped electrons would suggest extremely large defect density, which is not compatible with HRTEM, GXID and other experimental results. A possible explanation in terms of electrons trapped in in-gap Ti3d-O2p states at the interface has been proposed. The origin of these interface states is unclear and not necessarily related to extrinsic distribution of defects in STO.

However, some, very robust, experimental results agrees with the electronic reconstruction scenario: first, the system becomes conducting only at a critical thickness closely matching the expectations; the LAO/STO interface shows an orbital reconstruction, a distinctive feature of this system, which split the 3d-bands such that $3d_{xy}$ orbitals are the lowest in energy [62]; the structural evolution of the LAO films as function of the thickness closely follows the expected trend assuming the presence of electrostriction effects [49]. However, structural and XAS spectroscopic data [37], core-level photoemission spectroscopy [50], second har-monic generation [83] suggest an orbital reconstruction below the critical thickness.

It seems that a unifying explanation of the whole data is now still missing, however the polar discontinuity idea is so important and robust that other interfaces have been realized employing other polar-oxides and STO. One of the probably most puzzling aspects is that, at the moment, STO seems to be quite a unique and fundamental ingredient in the formation of the q2DES. However, from surface science we are learning about other potential materials, which can replace STO, like tantalates ($KTaO_3$ [84]), forming at their surfaces q2DES's.

Concerning the ground state properties of the q2DES, part of the controversies (superconductivity vs. magnetism) can be probably understood by taking into account the role of defects (presumably changing from sample to sample) and in particular of oxygen vacancies, as suggested by the study of ref. [61]. From these data it emerges a scenario for magnetism at the LAO/STO interface in the presence of extrinsic defects. Other point of views are magnetism appearing only at large carrier densities, normally achievable by introducing oxygen vacancies or by very strong electrostatic gating, or an intrinsic magnetism associated to the localized electrons necessary to solve the polar catastrophe. However, the study on annealed LAO/STO samples suggest that localized electrons are not in purely $Ti^{3+}$ states, but in in-gap states quite far from the Fermi level.

The new research trends in this field are concentrated on several directions: the realization of LAO/STO epitaxial films on different substrates, and even on silicon, with the aim to control and to make better the properties of the STO layers; the realization of complex structures, like quantum wells or interacting q2DES; the realization of oxide systems based on materials different than STO, like other titanates, to engineer both the carrier density and the mobility; the realization of nanostructures and nano-devices [85, 86].

Moreover, few groups are looking at δ-doping with other functional oxides in order to induce robust ferromagnetism or to enhance the spin-orbit coupling. Some attempts in these directions include the growth of $LaAlO_3/EuTiO_3/SrTiO_3$ [87], showing low T ferromagnetism and 2D-conductivity, $LaTiO_3/LaCrO_3/SrTiO_3$ systems [88], the deposition of ferromagnetic polar layer on STO and other oxides, or the use of other $3d^0$ materials like $KTaO_3$. Similarly, the atomic control achieved in the growth of these oxides has now provided ways to realize heterostructures where polar discontinuity can be engineered and modified using (110) and (111) interfaces [89, 90]. In the latter case, theoretical papers have suggested interesting topological protected states with some analogies with topological insulators [91].

Thus the future of 2D-oxide artificial systems is very bright and promise to provide unforeseen opportunities in technologies and science.

## Acknowledgments

The author is grateful to G. Ghiringhelli, F. Miletto Granozio, N. B. Brookes, J-M. Triscone, J. Mannhart, S. Gariglio, C. Piamonteze, X. Torrelles, S. Rusponi, D. Marrè, and R. Felici for the fruitful discussion and continuous collaboration on the subject. We also acknowledge E. di Gennaro, A. Brinkman, M.P. Warusawithana, M. Sing for granting permission about reproduction of some figures and for exchange of ideas. This research was funded by the European Union FP7 Program MPNS COST Action MP1308- TO-BE and from the Italian MIUR FIRB project Grant no. RBAP115AYN.

## References

[1] M.B. Salamon and M. Jaime, Rev. Mod. Phys. 73, 583 (2001).
[2] E. Fradkin and S.A. Kivelson, Nature Physics 8, 864 (2012).
[3] J. Mannhart and D.G. Schlom, Science 327, 1607 (2010).
[4] A. Ohtomo, D.A. Muller, J.L. Grazul, and H.Y. Hwang, Nature 419, 378 (2002).
[5] M. Imada, A. Fujimori, and Y. Tokura, Rev. Mod. Phys. 70, 1039 (1998).
[6] A. Ohtomo and H.Y. Hwang, Nature 427, 423 (2004).
[7] M. Huijben, G. Koster, M.K. Kruize, S. Wenderich, J. Verbeeck, S. Bals, E. Slooten, B. Shi, H.J.A. Molegraaf, J.E. Kleibeuker, S. van Aert, J.B. Goedkoop, A. Brinkman, D.H.A. Blank, M.S. Golden, G. Van Tendeloo, H. Hilgenkamp, and G. Rijnders, Adv. Funct. Mater. 23, 5240 (2013).
[8] N. Reyren, S. Thiel, A.D. Caviglia, L.F. Kourkoutis, G. Hammerl, C. Richter, C.W. Schneider, T. Kopp, A.S. Ruetschi, D. Jaccard, M. Gabay, D.A. Muller, J.M. Triscone, and J. Mannhart, Science 317, 1196 (2007).
[9] A.D. Caviglia, S. Gariglio, N. Reyren, D. Jaccard, T. Schneider, M. Gabay, S. Thiel, G. Hammerl, J. Mannhart, and J.M. Triscone, Nature 456, 624 (2008).
[10] S. Thiel, G. Hammerl, A. Schmehl, C.W. Schneider, and J. Mannhart, Science 313, 1942 (2006).
[11] A.D. Caviglia, M. Gabay, S. Gariglio, N. Reyren, C. Cancellieri, and J.M. Triscone, Phys. Rev. Lett. 104, 126803 (2010).
[12] M.A. Hein, Supercond. Sci. Technol. 10, 867 (1997).
[13] J.H. Song, T. Susaki, and H.Y. Hwang, Adv. Mater. **20**, 2528 (2008).
[14] M. Salluzzo, C. Aruta, I. Maggio-Aprile, Ø. Fischer, S. Bals, and J. Zegenhagen, Phys. Stat. Sol. (a) **186**, 339 (2001);
[15] M. Salluzzo, G. De Luca, D. Marrè, M. Putti, M. Tropeano, U. Scotti di Uccio, and R. Vaglio, Phys. Rev. B **72**, 134521 (2005).
[16] R. Scherwitzl, S. Gariglio, M. Gabay, P. Zubko, M. Gibert, and J.M. Triscone, Phys. Rev. Lett. **106**, 246403 (2011).
[17] K. A. Müller and H. Burkard, Phys. Rev. B 19, 3593 (1979).
[18] J.H. Haeni, P. Irvin, W. Chang, R. Uecker, P. Reiche, Y.L. Li, S. Choudhury, W. Tian, M.E. Hawley, and B. Craigo, Nature 430, 758 (2004).
[19] S. G. Lim, S. Kriventsov, T. N. Jackson, J. H. Haeni, D. G. Schlom, A. M. Balbashov, R. Uecker, P. Reiche, J. L. Freeouf, and G. Lucovsky, J. Appl. Phys. 91, 4500 (2002).
[20] M. Cardona, Phys. Rev. 140, 651 (1965).

[21] A. Spinelli, M. A. Torija, C. Liu, C. Jan, and C. Leighton, Phys. Rev. B 81, 155110 (2010).
[22] J. Appel, Soft-Mode Superconductivity in SrTiO$_3$x, Phys. Rev. 180, 508 (1969).
[23] X. Lin, Z. Zhu, B. Fauque, and K. Behnia, Phys. Rev. X 3, 021002 (2013).
[24] M. Kawasaki, K. Takahashi, T. Maeda, R. Tsuchiya, M. Shinohara, O. Ishiyama, T. Yonezawa, M. Yoshimoto, H. Koinuma, Science 266 (1994) 1540.
[25] G. Koster, B.L. Kropman, G.J. Rijnders, D.H. Blank, and H. Rogalla, Appl. Phys. Lett. **73**, 2920 (1998).
[26] A. Fragneto, G.M. De Luca, R. Di Capua, U. Scotti di Uccio, M. Salluzzo, X. Torrelles, T.-L. Lee, and J. Zegenhagen, Appl. Phys. Lett. **91**, 101910 (2007).
[27] T. Kubo and H. Nozoye, Surface Science **542**, 177 (2003).
[28] R. Eglitis and D. Vanderbilt, Phys. Rev. B **77**, 195408 (2008).
[29] P.W. Tasker, J. Phys. C: Solid State Phys. 12, 4977–4984. (1979)
[30] R. Francis, S. Moss, and A. Jacobson, Phys. Rev. B **64**, 235425 (2001).
[31] H. Lehnert, H. Boysen, P. Dreier, and Y. Yu, Z- Kristallogr. **215**, 145 (2000).
[32] G.J. Rijnders, G. Koster, D.H. Blank, and H. Rogalla, Appl. Phys. Lett. **70**, 1888 (1997).
[33] M. Huijben, A. Brinkman, G. Koster, G. Rijnders, H. Hilgenkamp, and D.H.A. Blank, Adv. Mater. **21**, 1665 (2009).
[34] T. Ohnishi, M. Lippmaa, T. Yamamoto, S. Meguro, and H. Koinuma, Appl. Phys. Lett. **87**, 241919 (2005).
[35] S. Wicklein, A. Sambri, S. Amoruso, X. Wang, R. Bruzzese, A. Koehl, and R. Dittmann, Appl. Phys. Lett. **101**, 131601 (2012).
[36] A. Sambri, D.V. Cristensen, F. Trier, Y.Z. Chen, S. Amoruso, N. Pryds, R. Bruzzese, and X. Wang, Appl. Phys. Lett. **100**, 231605 (2012).
[37] M. Salluzzo, S. Gariglio, X. Torrelles, Z. Ristic, R. Di Capua, J. Drnec, M.M. Sala, G. Ghiringhelli, R. Felici, and N.B. Brookes, Adv. Mater. **25**, 2333 (2013).
[38] M.P. Warusawithana, C. Richter, J.A. Mundy, P. Roy, J. Ludwig, S. Paetel, T. Heeg, A.A. Pawlicki, L.F. Kourkoutis, M. Zheng, M. Lee, B. Mulcahy, W. Zander, Y. Zhu, J. Schubert, J.N. Eckstein, D.A. Muller, C.S. Hellberg, J. Mannhart, and D.G. Schlom, *Nat. Commun.* 4:2351 doi: 10.1038/ncomms3351 (2013).
[39] G. Herranz, M. Basletic, M. Bibes, C. Carrétéro, E. Tafra, E. Jacquet, K. Bouzehouane, C. Deranlot, A. Hamzić, J.M. Broto, A. Barthélémy, and A. Fert, Phys. Rev. Lett. **98**, 216803 (2007).
[40] C. Aruta, S. Amoruso, G. Ausanio, R. Bruzzese, E. Di Gennaro, M. Lanzano, F.M. Granozio, M. Riaz, A. Sambri, and U.S. di Uccio, Appl. Phys. Lett. **101**, 031602 (2012).
[41] A. Brinkman, M. Huijben, M. van Zalk, J. Huijben, U. Zeitler, J.C. Maan, W.G. van der Wiel, G. Rijnders, D.H.A. Blank, and H. Hilgenkamp, Nature Materials **6**, 493 (2007).
[42] C. Aruta, S. Amoruso, G. Ausanio, R. Bruzzese, E. Di Gennaro, M. Lanzano, F.M. Granozio, M. Riaz, A. Sambri, and U.S. di Uccio, Appl. Phys. Lett. **101**, 031602 (2012).
[43] E. Di Gennaro, U.S. di Uccio, C. Aruta, C. Cantoni, A. Gadaleta, A.R. Lupini, D. Maccariello, D. Marré, I. Pallecchi, D. Paparo, P. Perna, M. Riaz, and F.M. Granozio, Advanced Optical Materials **1**, 834 (2013).
[44] Y.Z. Chen, N. Bovet, F. Trier, D.V. Christensen, F.M. Qu, N.H. Andersen, T. Kasama, W. Zhang, R. Giraud, J. Dufouleur, T.S. Jespersen, J.R. Sun, A. Smith, J. Nygård, L. Lu, B. Büchner, B.G. Shen, S. Linderoth, and N. Pryds, Nature Communications **4**, 1371 (2013).
[45] R. Pentcheva and W. Pickett, Phys. Rev. Lett. **102**, 107602 (2009).
[46] At the moment there are no indications that the (001) LaAlO$_3$ can be at equilibrium LAO terminated. According to ref. [30] the stable termination of LaAlO$_3$ single crystals is AlO$_x$.
[47] A. Janotti, L. Bjaalie, L. Gordon, and C.G. Van de Walle, Phys. Rev. B **86**, 241108 (2012).
[48] S.A. Pauli, S.J. Leake, B. Delley, M. Björck, C.W. Schneider, C.M. Schlepütz, D. Martoccia, S. Paetel, J. Mannhart, and P.R. Willmott, Phys. Rev. Lett. **106**, 036101 (2011).

[49] C. Cancellieri, D. Fontaine, S. Gariglio, N. Reyren, A.D. Caviglia, A. Fête, S.J. Leake, S.A. Pauli, P.R. Willmott, M. Stengel, P. Ghosez, and J.M. Triscone, Phys. Rev. Lett. **107**, 056102 (2011).

[50] S. A. Chambers, M. H. Englehard, V. Shutthanandan, Z. Zhu, T. C. Droubay, T. Feng, H. D. Lee, T. Gustafsson, E. Garfunkel, A. Shah, J. M. Zuo, and Q. M. Ramasse, Surf. Sci. Rep. 65, 317 (2010).

[51] M. Sing, G. Berner, K. Goß, A. Müller, A. Ruff, A. Wetscherek, S. Thiel, J. Mannhart, S. Pauli, C. Schneider, P. Willmott, M. Gorgoi, F. Schäfers, and R. Claessen, Phys. Rev. Lett. **102**, 176805 (2009).

[52] Y. Chen, N. Pryds, J.E. Kleibeuker, G. Koster, J. Sun, E. Stamate, B. Shen, G. Rijnders, and S. Linderoth, Nano Lett. 11, 3774 (2011).

[53] Z.Q. Liu, C.J. Li, W.M. Lü, X.H. Huang, Z. Huang, S.W. Zeng, X.P. Qiu, L.S. Huang, A. Annadi, J.S. Chen, J.M.D. Coey, T. Venkatesan, Ariando, Phys. Rev. X **3**, 021010 (2013).

[54] N. Nakagawa, H.Y. Hwang, and D.A. Muller, Nature Materials 5, 204 (2006).

[55] P. Willmott, S. Pauli, R. Herger, C. Schlepütz, D. Martoccia, B. Patterson, B. Delley, R. Clarke, D. Kumah, C. Cionca, and Y. Yacoby, Phys. Rev. Lett. 99, 155502 (2007).

[56] N.C. Bristowe, P. Ghosez, P.B. Littlewood, and E. Artacho, *J. Phys.: Condens. Matter* **26**, 143201 (2014).

[57] L. Yu and A. Zunger, arXiv cond-mat.mtrl-sci, arXiv:1402.0895 (2014).

[58] G. Berner, M. Sing, H. Fujiwara, A. Yasui, Y. Saitoh, A. Yamasaki, Y. Nishitani, A. Sekiyama, N. Pavlenko, and T. Kopp, Phys. Rev. Lett. **110**, 247601 (2013).

[59] C. Cen, S. Thiel, G. Hammerl, C.W. Schneider, K.E. Andersen, C.S. Hellberg, J. Mannhart, and J. Levy, Nature Materials **7**, 298 (2008).

[60] F. De Groot, Journal of Electron Spectroscopy and Related Phenomena **67**, 529 (1994).

[61] M. Salluzzo, S. Gariglio, D. Stornaiuolo, V. Sessi, S. Rusponi, C. Piamonteze, G.M. De Luca, M. Minola, D. Marrè, A. Gadaleta, H. Brune, F. Nolting, N.B. Brookes, and G. Ghiringhelli, Phys. Rev. Lett. **111**, 087204 (2013).

[62] M. Salluzzo, J. Cezar, N. Brookes, V. Bisogni, G. De Luca, C. Richter, S. Thiel, J. Mannhart, M. Huijben, A. Brinkman, G. Rijnders, and G. Ghiringhelli, Phys. Rev. Lett. **102**, 166804 (2009).

[63] In the ionic formal configuration, oxygen is in the $O^{2-}$ valence state, with fully occupancy of the 2p orbitals ($2p^6$). One may wonder why 1s-2p process is allowed in the solid. The reason is that in a solid oxygen and neighbour cations are always characterized by some degree of covalence bonding (the bonding is not purely ionic), and as consequence some of the 2p electrons are not exactly localized in the 2p states, but shared among the other cations and in particular with Ti-3d states. This is one of the reasons why the assignment of a formal static valence to an ion in a solid is in-general quite arbitrary, and may generate confusion. In particular, $Ti^{3+}$ and $Ti^{4+}$ valence states are probably close to the effective valence only in the limiting cases of $SrTiO_3$ ($Ti^{4+}$ $3d^0$) and $LaTiO_3$ ($Ti^{3+}$ $3d^1$). Still, a more appropriate definition of the charge of an atom in ionic crystal is necessary

[64] Z. Ristic, R. Di Capua, G.M. De Luca, F. Chiarella, G. Ghiringhelli, J.C. Cezar, N.B. Brookes, C. Richter, J. Mannhart, and M. Salluzzo, Europhys. Lett. **93**, 17004 (2011).

[65] C. Cancellieri, M.L. Reinle-Schmitt, M. Kobayashi, V.N. Strocov, P.R. Willmott, D. Fontaine, P. Ghosez, A. Filippetti, P. Delugas, and V. Fiorentini, Phys. Rev. B **89**, 121412 (2014).

[66] A.F. Santander-Syro, O. Copie, T. Kondo, F. Fortuna, S. Pailhès, R. Weht, X.G. Qiu, F. Bertran, A. Nicolaou, A. Taleb-Ibrahimi, P. Le Fèvre, G. Herranz, M. Bibes, N. Reyren, Y. Apertet, P. Lecoeur, A. Barthélémy, and M.J. Rozenberg, Nature **469**, 189 (2010).

[67] N.C. Plumb, M. Salluzzo, E. Razzoli, M. Månsson, M. Falub, J. Krempasky, C.E. Matt, J. Chang, M. Schulte, J. Braun, H. Ebert, J. Minár, B. Delley, K.J. Zhou, T. Schmitt, M. Shi, J. Mesot, L. Patthey, and M. Radovic, Phys. Rev. Lett. 113, 086801 (2014).

[68] R. Pentcheva and W. Pickett, Phys. Rev. B **74**, 035112 (2006).

[69] P. Delugas, A. Filippetti, V. Fiorentini, D.I. Bilc, D. Fontaine, and P. Ghosez, Phys. Rev. Lett. **106**, 166807 (2011).

[70] M. Breitschaft, V. Tinkl, N. Pavlenko, S. Paetel, C. Richter, J.R. Kirtley, Y.C. Liao, G. Hammerl, V. Eyert, T. Kopp, and J. Mannhart, Phys. Rev. B **81**, 153414 (2010).

[71] C. Richter, H. Boschker, W. Dietsche, E. Fillis-Tsirakis, R. Jany, F. Loder, L.F. Kourkoutis, D.A. Muller, J.R. Kirtley, C.W. Schneider, and J. Mannhart, Nature **502**, 528 (2013).

[72] Z. Ristic, R. Di Capua, F. Chiarella, G.M. De Luca, I. Maggio-Aprile, M. Radović, and M. Salluzzo, Phys. Rev. B **86**, 045127 (2012).

[73] Y. Ishida, R. Eguchi, M. Matsunami, K. Horiba, M. Taguchi, A. Chainani, Y. Senba, H. Ohashi, H. Ohta, and S. Shin, Phys. Rev. Lett. **100**, 056401 (2008).

[74] Ariando, X. Wang, G. Baskaran, Z.Q. Liu, J. Huijben, J.B. Yi, A. Annadi, A.R. Barman, A. Rusydi, S. Dhar, Y.P. Feng, J. Ding, H. Hilgenkamp, and T. Venkatesan, Nature Communications **2**, 188 (2011).

[75] L. Li, C. Richter, J. Mannhart, and R.C. Ashoori, Nature Physics **7**, 762 (2011).

[76] J.A. Bert, B. Kalisky, C. Bell, M. Kim, Y. Hikita, H.Y. Hwang, and K.A. Moler, Nature Physics **7**, 767 (2011).

[77] M.R. Fitzsimmons, N.W. Hengartner, S. Singh, M. Zhernenkov, F.Y. Bruno, J. Santamaria, A. Brinkman, M. Huijben, H.J.A. Molegraaf, J. de la Venta, and I.K. Schuller, Phys. Rev. Lett. **107**, 217201 (2011).

[78] Z. Salman, O. Ofer, M. Radović, H. Hao, M. Ben Shalom, K.H. Chow, Y. Dagan, M.D. Hossain, C.D.P. Levy, W.A. MacFarlane, G.M. Morris, L. Patthey, M.R. Pearson, H. Saadaoui, T. Schmitt, D. Wang, and R.F. Kiefl, Phys. Rev. Lett. **109**, 257207 (2012).

[79] P. Gambardella, S. Rusponi, M. Veronese, S. S. Dhesi, C. Grazioli, A. Dallmeyer, I. Cabria, R. Zeller, P. H. Dederichs, K. Kern, C. Carbone, and H. Brune, Science 300, 1130 (2003).

[80] J.S. Lee, Y.W. Xie, H.K. Sato, C. Bell, Y. Hikita, H.Y. Hwang, and C.C. Kao, Nature Materials **12**, 703 (2013).

[81] E. Stavitski and F.M.F. de Groot, Micron 41, 687 (2010).

[82] S. Banerjee, O. Erten, and M. Randeria, Nature Physics 9, 626 (2013).

[83] A. Savoia, D. Paparo, P. Perna, Z. Ristic, M. Salluzzo, F. Miletto Granozio, U. Scotti di Uccio, C. Richter, S. Thiel, J. Mannhart, and L. Marrucci, Phys. Rev. B **80**, 075110 (2009).

[84] A.F. Santander-Syro, C. Bareille, F. Fortuna, O. Copie, M. Gabay, F. Bertran, A. Taleb-Ibrahimi, P. Le Fèvre, G. Herranz, N. Reyren, M. Bibes, A. Barthélémy, P. Lecoeur, J. Guevara, and M.J. Rozenberg, Phys. Rev. B **86**, 121107 (2012).

[85][1] D. Stornaiuolo, et al., Appl. Phys. Lett. **101**, 222601 (2012).

[86][1] R. Jany, et al., Adv. Mater. Interfaces **1**, 1300031 (2013).

[87] G.M. De Luca, R. Di Capua, E. Di Gennaro, F.M. Granozio, D. Stornaiuolo, M. Salluzzo, A. Gadaleta, I. Pallecchi, D. Marrè, C. Piamonteze, M. Radović, Z. Ristic, and S. Rusponi, Phys. Rev. B **89**, 224413 (2014).

[88] S. Das, A. Rastogi, L. Wu, J.-C. Zheng, Z. Hossain, Y. Zhu, and R.C. Budhani, Phys. Rev. B **90**, 081107 (2014).

[89] A. Annadi, Q. Zhang, X.R. Wang, N. Tuzla, K. Gopinadhan, W.M.L. uuml, A.R. Barman, Z.Q. Liu, A. Srivastava, S. Saha, Y.L. Zhao, S.W. Zeng, S. Dhar, E. Olsson, B. Gu, S. Yunoki, S. Maekawa, H. Hilgenkamp, T. Venkatesan, Ariando, Nature Communications **4**, 1838 (2013).

[90] G. Herranz, F. Sánchez, N. Dix, M. Scigaj, and J. Fontcuberta, Nature Communications 2, (2012).

[91] D. Doennig, W.E. Pickett, and R. Pentcheva, Phys. Rev. Lett. **111**, 126804 (2013).